\documentclass[12pt]{article}

\usepackage{natbib} 
\usepackage{color} 
\usepackage[table]{xcolor} 
\usepackage{paralist}
\usepackage{geometry}
\usepackage{graphicx,graphics}
\usepackage{amsmath}	
\usepackage{amssymb}	
\usepackage{newtxtext,newtxmath}
\usepackage[T1]{fontenc}
\usepackage{ae,aecompl}
\usepackage{epsfig}   
\usepackage{microtype}
\usepackage[utf8]{inputenc}
\usepackage{enumitem}
\usepackage{ragged2e}
\usepackage{times}
\newlist{thematic}{itemize}{8}
\setlist[thematic]{label=$\square$}
\usepackage{pifont}
\usepackage{enumitem} 

\newcommand{\specialcell}[2][c]{%
  \begin{tabular}[#1]{@{}c@{}}#2\end{tabular}}

\def\capfontcom{\normalsize\color{gray!60!black}}
\long\def\@makecaption#1#2{%
  \vskip\abovecaptionskip
  {\centering
    \begin{minipage}{0.999\linewidth}
      \sbox\@tempboxa{{\sffamily\bfseries\small #1}\capfontcom\,--- #2}%
      \ifdim \wd\@tempboxa >\hsize
      {{\sffamily\bfseries\small #1}\capfontcom\,--- #2}
      \else
      \global \@minipagefalse
      \hb@xt@\hsize{\hfil\box\@tempboxa\hfil}%
      \fi
      \end{minipage}\par
      }
  \vskip\belowcaptionskip}

\definecolor{tablealt}{rgb}{0.77,0.85,1.0}

\def\aj{{AJ}}                   
\def\araa{{ARA\&A}}          
\def\apj{{ApJ}}                 
\def\apjl{{ApJ}}

\def\aap{ {A\&A}}

\def\mnras{ {MNRAS}}

\def\pasj{ {PASJ}}

\newcommand{\be}{\begin{equation}}
\newcommand{\ee}{\end{equation}}

\newcommand{\gtsima}{$\; \buildrel > \over \sim \;$}
\newcommand{\ltsima}{$\; \buildrel < \over \sim \;$}
\newcommand{\prosima}{$\; \buildrel \propto \over \sim \;$}
\newcommand{\gsim}{\lower.5ex\hbox{\gtsima}}
\newcommand{\lsim}{\lower.5ex\hbox{\ltsima}}
\newcommand{\simgt}{\lower.5ex\hbox{\gtsima}}
\newcommand{\simlt}{\lower.5ex\hbox{\ltsima}}
\newcommand{\simpr}{\lower.5ex\hbox{\prosima}}

\newcommand{\arcmin}{$^{\prime}$}
\newcommand{\arcsec}{$^{\prime\prime}$}

\begin{document}

\pagenumbering{roman}

\raggedright
\huge
Astro2020 Science White Paper \linebreak

How does dust escape galaxies? \linebreak
\normalsize

\noindent \textbf{Thematic Areas:} \hspace*{60pt} $\square$ Planetary Systems \hspace*{10pt} $\square$ Star and Planet Formation \hspace*{20pt}\linebreak
$\square$ Formation and Evolution of Compact Objects \hspace*{31pt} $\square$ Cosmology and Fundamental Physics \linebreak
  $\square$  Stars and Stellar Evolution \hspace*{1pt} $\square$ Resolved Stellar Populations and their Environments \hspace*{40pt} \linebreak
  $\boxtimes$    Galaxy Evolution   \hspace*{45pt} $\square$             Multi-Messenger Astronomy and Astrophysics \hspace*{65pt} \linebreak
  
\textbf{Principal Author:}

Name: Edmund Hodges-Kluck
 \linebreak						
Institution:  University of Maryland/NASA Goddard Space Flight Center
 \linebreak
Email: edmund.hodges-kluck@nasa.gov
 \linebreak
 
\textbf{Co-authors:}  \linebreak
L\'ia Corrales (University of Michigan)\\
Sylvain Veilleux (University of Maryland)\\
Joel N. Bregman (University of Michigan)\\
Jiangtao Li (University of Michigan)\\
Marcio Mel\'endez (Space Telescope Science Institute)\\

\justify

\textbf{Abstract:} Whenever gas is blown out of a galaxy, chances are that it contains some cosmic dust. This dust is an important part of the metals budget for the circumgalactic and intergalactic media, and traces the outflow and stripping history of the galaxy. The dust is also interesting in its own right, as dust plays an essential role in many astrophysical processes. We have only begun to learn about circumgalactic dust, and in particular we do not know how (and when) it escapes its host galaxy. Here we describe the prospects for measuring the dust mass and properties around many individual galaxies, which will form the basis for understanding how the dust got there. 

\pagebreak

\setcounter{page}{1}
\pagenumbering{arabic}

\noindent \textbf{\sf \large 1. Circumgalactic Dust}
\medskip

The circumgalactic medium (CGM) has recently become a hot topic---rightfully so, as it contains much (or most) of the non-dark matter associated with galaxies from $z>10$ to $z=0$ \citep{Tumlinson2017}. The CGM also contains a large amount of dust \citep[by some estimates, more than in the galaxy itself;][]{Menard2010}. Whereas the fraction of the CGM that has ever cycled through a galaxy remains controversial, 
\textbf{every dust grain in the CGM was born in a galaxy.}

Circumgalactic dust (CGD) is thus an important tracer of the integrated outflow and gas-stripping history of the galaxy, especially as it is long-lived in the tenuous CGM \citep{Ferrara1991}. Dust can also \textit{drive} outflows through radiation pressure on grains \citep{Murray2005}. Once outside of a galaxy, CGD can cool or heat the CGM \citep{Montier2004}. 

\begin{figure}[ht]
\centering
\includegraphics[width=0.95\textwidth]{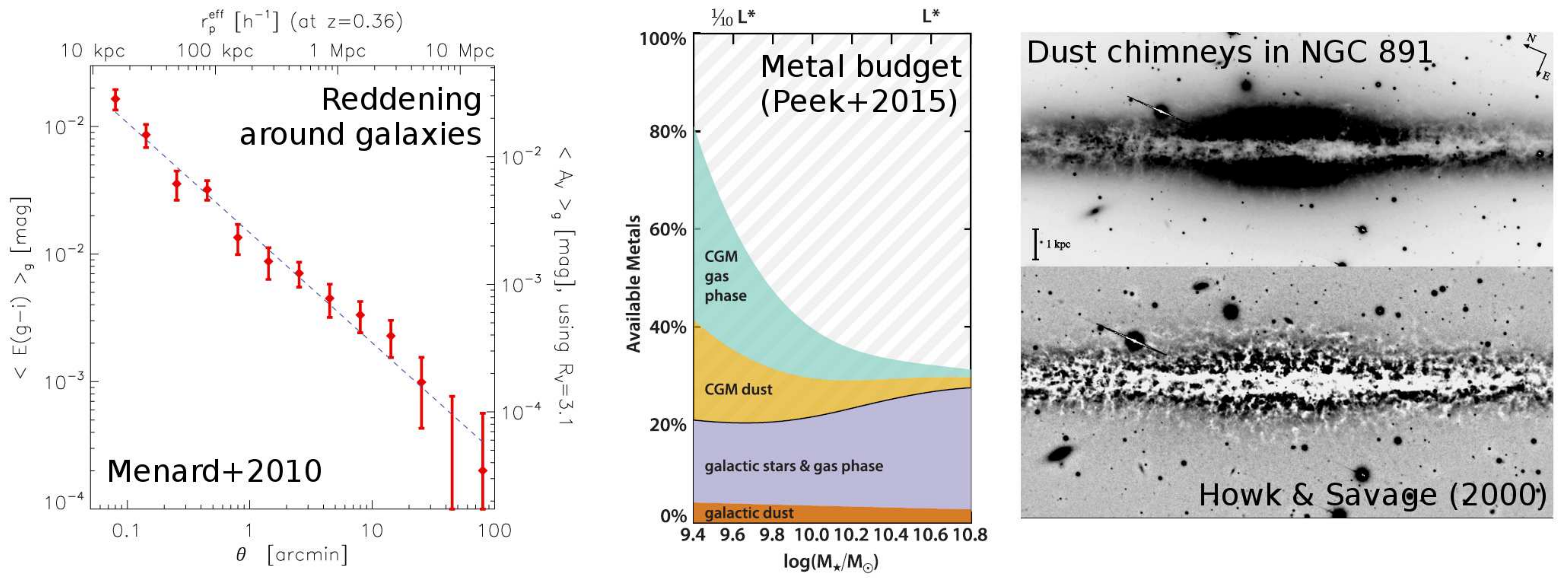}
\caption
{\small \textbf{Left:} Quasar reddening by dust is seen around galaxies to beyond $R_{\text{vir}}$ (adapted from \citealt{Menard2010}). \textbf{Center:} Reddening implies a dust mass, and around low-mass galaxies extragalactic dust contains 10-20\% of the metals (adapted from \citealt{Peek2015}). \textbf{Right:} Vertical dust filaments in nearby galaxies show how supernova-blown bubbles break out of galaxies (adapted from \citealt{Howk2000}). 
\label{figure:extinction}
}
\end{figure}

There is a lot of CGD. Measurements of quasar reddening around galaxies at $z\approx 0.3$ revealed dust to beyond the virial radius (Figure~\ref{figure:extinction}), with a total mass exceeding that in the host galaxies \citep{Menard2010}, and CGD accounts for up to 20\% of the metal budget around small galaxies \citep{Peek2015}. The total mass remains uncertain because many sightlines pierce extended gaseous disks around the optical galaxies \citep{Smith2016}, but several lines of evidence point to large CGD masses. We would therefore like to know: 
\begin{itemize}[noitemsep,topsep=2mm,fullwidth,itemindent=0mm]
    \item How is CGD partitioned among each CGM phase?
    \item How does the CGD contribute to the metal budget as a function of galaxy mass?    
    \item When was CGD expelled/stripped from its natal galaxy, and how long does it live in the CGM? 
    \item Does CGD escape mostly via episodic winds, bubbling fountains,  starlight ``breezes,'' or stripping? 
\end{itemize}
Answering these questions requires measuring the mass, composition and size (hereafter, ``dust properties''), velocity, and spatial distribution of the CGD. Here we address how this can be done.

\medskip
\noindent \textbf{\sf \large 2. Where does the CGD live?}
\medskip

Reddening of quasars \citep{Menard2010} or background galaxies \citep[``standard crayons'';][]{Peek2015} around galaxies remains an important, and perhaps the best, tool to associate dust far from galaxies with phases of the CGM. For example, about half of the CGD may reside in Mg~{\sc ii} absorbers \citep{Menard2012}. However, one must account for contamination by extended gas disks, and future advances require careful selection of the sight lines (e.g., by azimuth).

This work was based on the SDSS (reaching 21-23~mag over 10,000~deg$^2$), but many more systems are available: the Dark Energy Survey reaches $m$=24~mag over 5,000~deg$^2$, Pan-STARRS reaches $m$=21-23~mag over 30,000~deg$^2$, LSST will reach $m$=27~mag over 30,000~deg$^2$, and J-PAS will cover the SDSS footprint with 54 filters from the optical to near-IR. The AllWISE catalog contains $\sim$1.4$\times$10$^6$~quasars \citep{Secrest2015}, and \textit{eROSITA} will discover many X-ray AGNs. By selecting samples following recent CGM studies, one can measure the average CGD properties as a function of galaxy type, and its association with different phases of gas. 

One can also constrain the amount of dust in the cool and warm phases by measuring depletion of O, Si, and Fe in CGM gas. Following surveys of the CGM with \textit{HST}/COS \citep{Borthakur2015,Heckman2017,Prochaska2017}, a sensitive UV spectrometer is needed to expand the sample to hundreds of galaxies; the CETUS Probe concept \citep{Danchi2018} meets these requirements and would make these measurements as part of its surveys. It may also be possible to detect truly intergalactic gas through X-ray scattering halos \citep{Corrales2012,Corrales2015} with a high angular resolution, sensitive telescope, such as the \textit{Lynx} HDXI \citep{Gaskin2018} or AXIS Probe concept \citep{Mushotzky2018}, and a bright background source. 

\medskip
\noindent \textbf{\sf \large 3. Normal Galaxies and CGD Detection Techniques}
\medskip

In galaxies without galaxy-scale winds, dust escapes the galaxy disk through momentum imparted from starlight (or cosmic rays), galactic fountains of hot gas powered by supernovae \citep{Bregman1980}, and tidal or ram-pressure stripping. The amount of dust entrained in the outflow depends on the feedback mechanism, outflow speed, grain size distribution, and composition. 

The efficiency of each channel at removing dust is unknown, and they are difficult to study in isolation. The velocities are likely insufficient to remove gas from the gravitational potential, so most of this dust may remain close to the galaxy. By measuring the properties of the inner CGD around hundreds of galaxies we can identify relations with respect to the star-formation rate, galaxy mass, galaxy type, and interaction stage or morphology. The cross-section is too low to use background AGNs, but the dust can be observed in emission, absorption, and scattering:

\begin{figure}[b]
\centering
\includegraphics[height=1.5in]{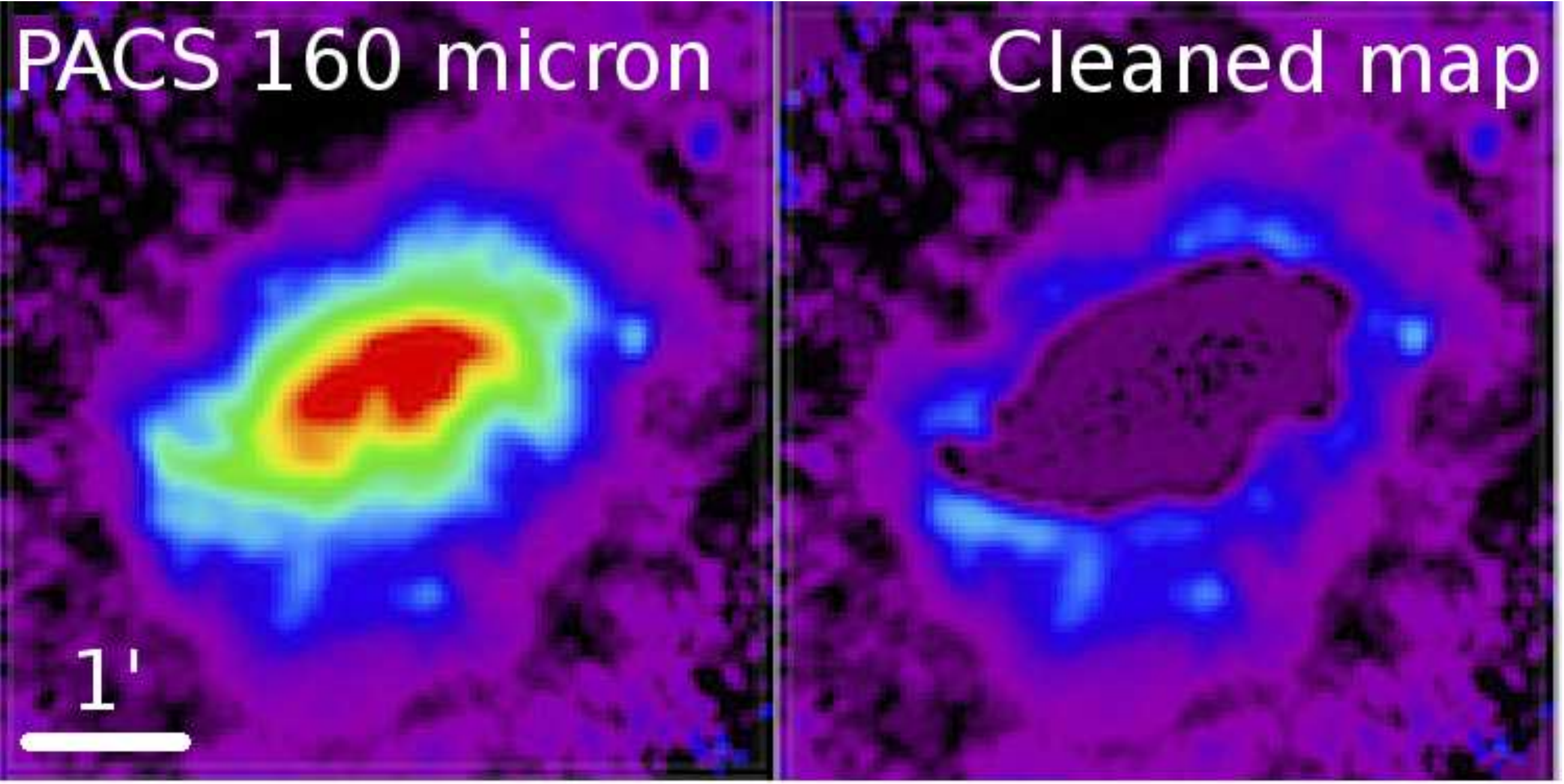}
\includegraphics[height=1.5in]{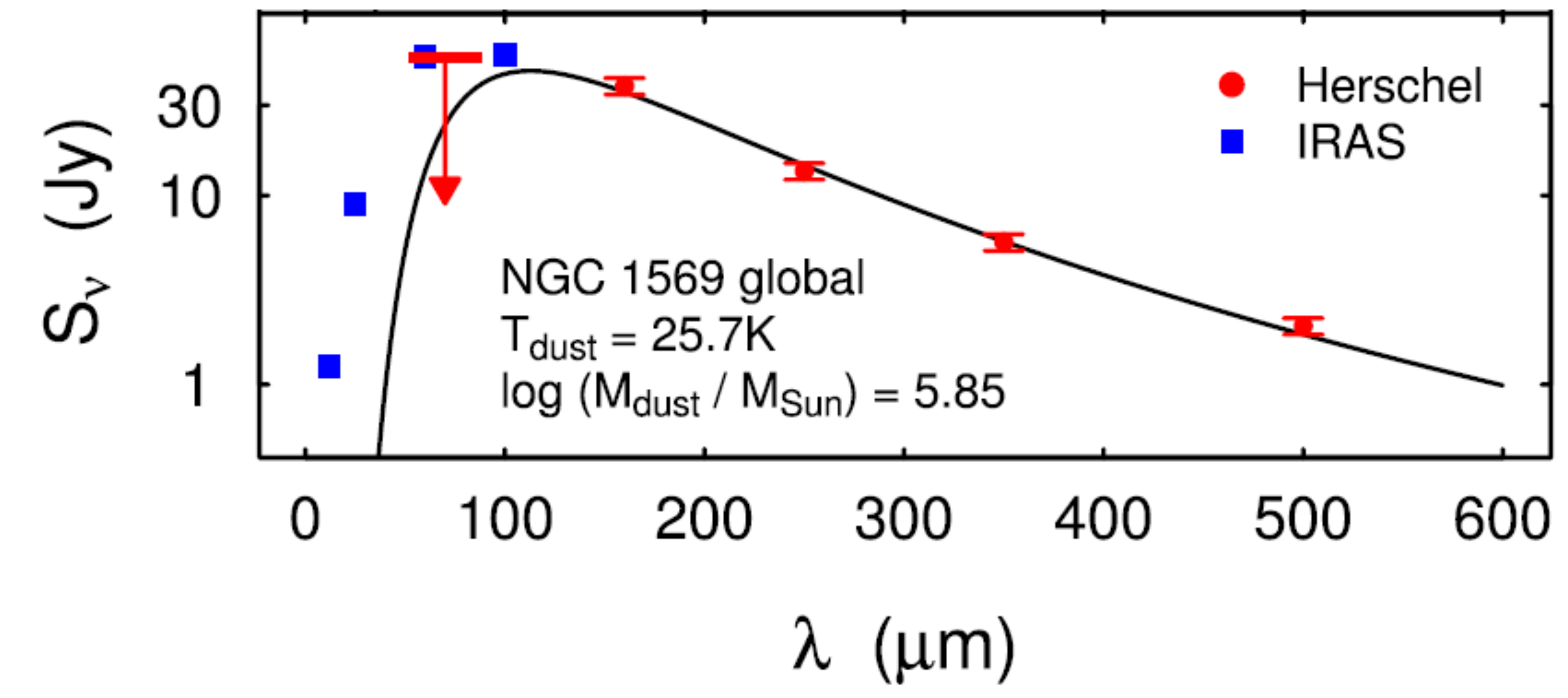}
\caption
{\small \textbf{Left:} \textit{Herschel} images can be cleaned to isolate extragalactic dust, as shown here for the dwarf starburst NGC~1569. \textbf{Right:} After doing this for several channels, SED fitting yields the dust temperature and mass (both panels adapted from \citealt{McCormick2013}). 
\label{figure:emission}
}
\end{figure}

\paragraph{Emission:} Dust processes radiation by absorbing UV or optical photons and radiating that energy in the infrared/sub-mm. The spectrum is represented by a modified blackbody, and if the temperature can be fixed by fitting the broadband spectral energy distribution (SED), the intensity yields the dust mass. Dust within several kpc of the galaxy disk is warm enough to detect in emission, and it has been detected around multiple galaxies with \textit{Herschel} (e.g., Figure~\ref{figure:emission}) by modeling and removing the galaxy image, including the PSF wings \citep{HK2016,McCormick2018}. 

The \textit{SOFIA}/HAWC$+$ camera has a similar sensitivity and resolution to \textit{Herschel} in the FIR, and could detect extraplanar dust within $\sim$25~Mpc, with several hours per galaxy. Sub-mm facilities (ALMA, NOEMA, and JCMT) can constrain the emissivity from the Rayleigh-Jeans tail, at the cost of long integrations at high frequency. A larger sample must include more distant galaxies and need 2-3x better angular resolution than \textit{Herschel}, which will also enable mapping of changes in the emissivity-temperature relationship and the dust-to-gas ratio in streams of tidally stripped gas \citep{Roussel2010,Melendez2015}. The \textit{Origins} concept meets the requirements for an ambitious survey, while the mid-infrared instruments on \textit{WFIRST} and \textit{JWST} will enable a systematic measurement of circumgalactic PAHs (polycyclic aromatic hydrocarbons).

\begin{figure}[t]
\vspace{0.3cm}
\centering
\includegraphics[height=2.2in]{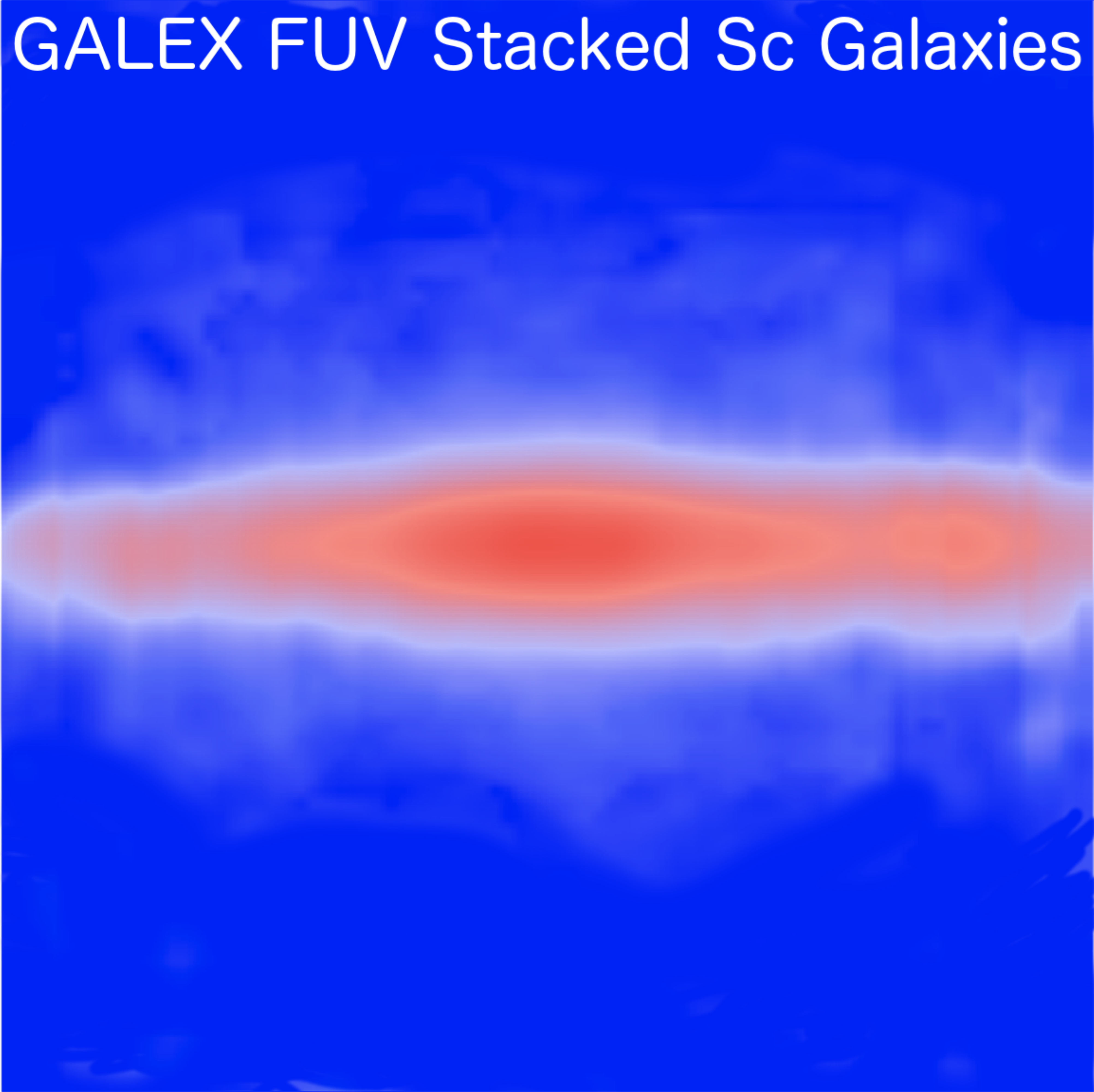}
\includegraphics[height=2.2in]{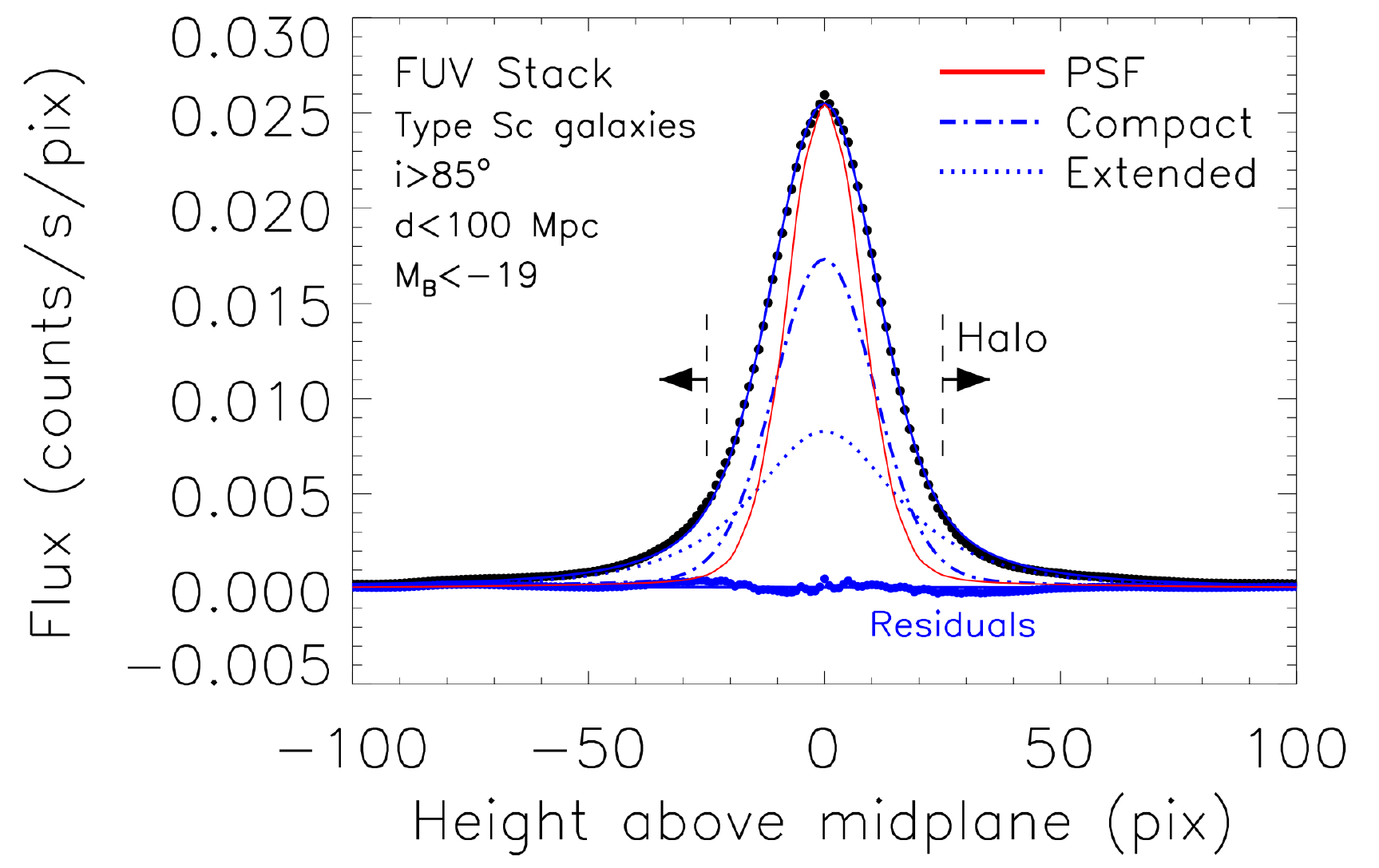}
\caption
{\small \textbf{Left:}  The stacked \textit{GALEX} FUV ($\lambda$1516\AA) image of edge-on type Sc galaxies within 100~Mpc and with $M_B < -19$~mag reveals diffuse halo light. \textbf{Right:} Much of the ``halo'' light is from the galaxy image (the wings of the PSF), but a single disk component cannot account for all of it (here we account for the full 2D PSF wings, but project to 1D). An extended component is needed, but note that the halo light is measured above several scale heights (dashed lines). All highly inclined, star-forming galaxy stacks have similar halos.
\label{figure:stacks}
}
\end{figure}

\paragraph{\sf Absorption:} Outflowing dust is seen in absorption against the background thick stellar disk in the walls of ``chimneys'' through which hot gas escapes \citep{Howk2000} (Figure~\ref{figure:extinction}). Most of this dust will not escape the galaxy, but it does provide information on the grain properties and outflow mass and energy budgets. Some filaments also indicate the sites of condensation of hot gas or its interaction with cooler gas \citep{Li2015}, and dust lanes in stripped material constrain the ``outflow'' rate. Further studies can be carried out with sensitive ground-based observations. 

\begin{figure}[t]
\centering
\includegraphics[height=2.1in]{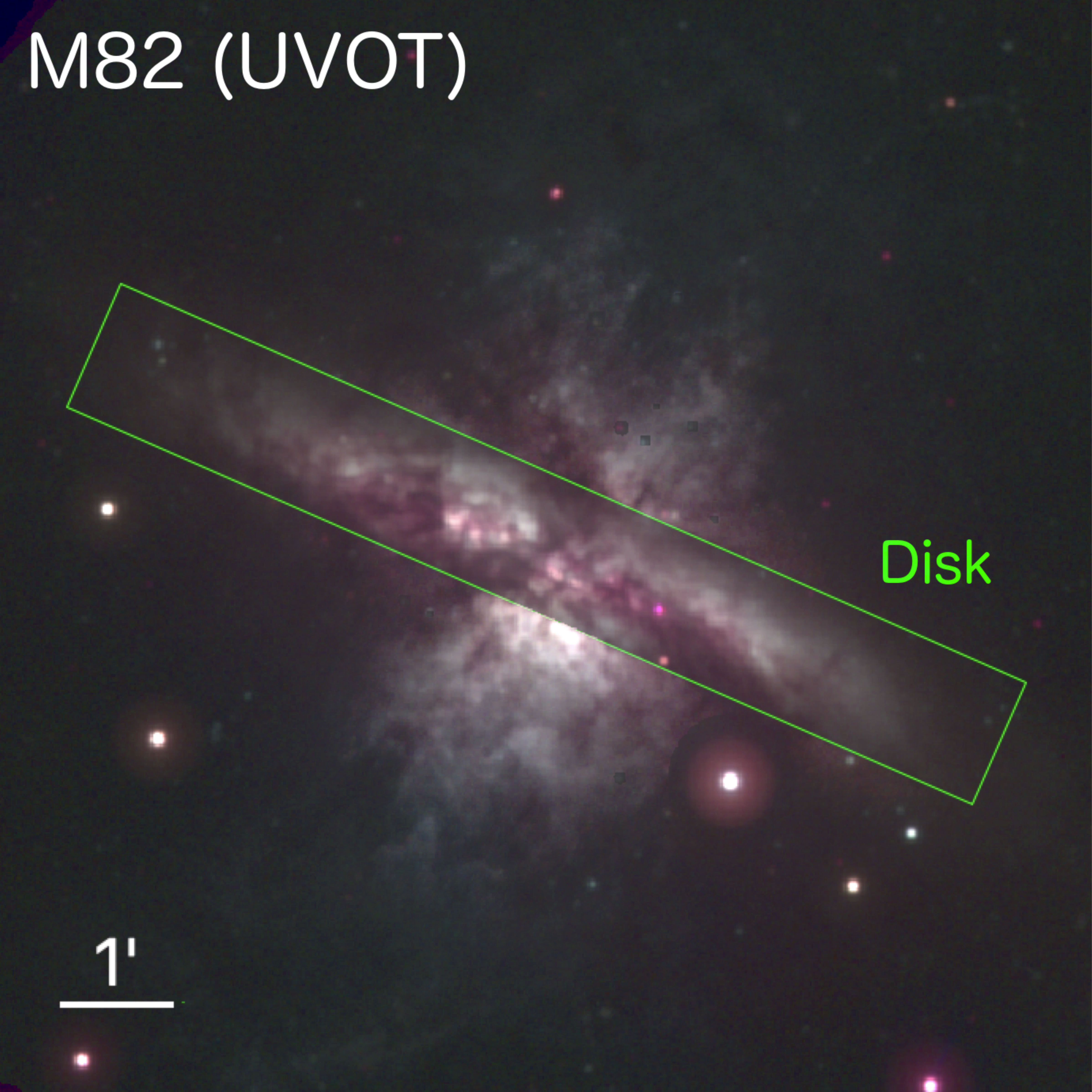}
\includegraphics[height=2.1in]{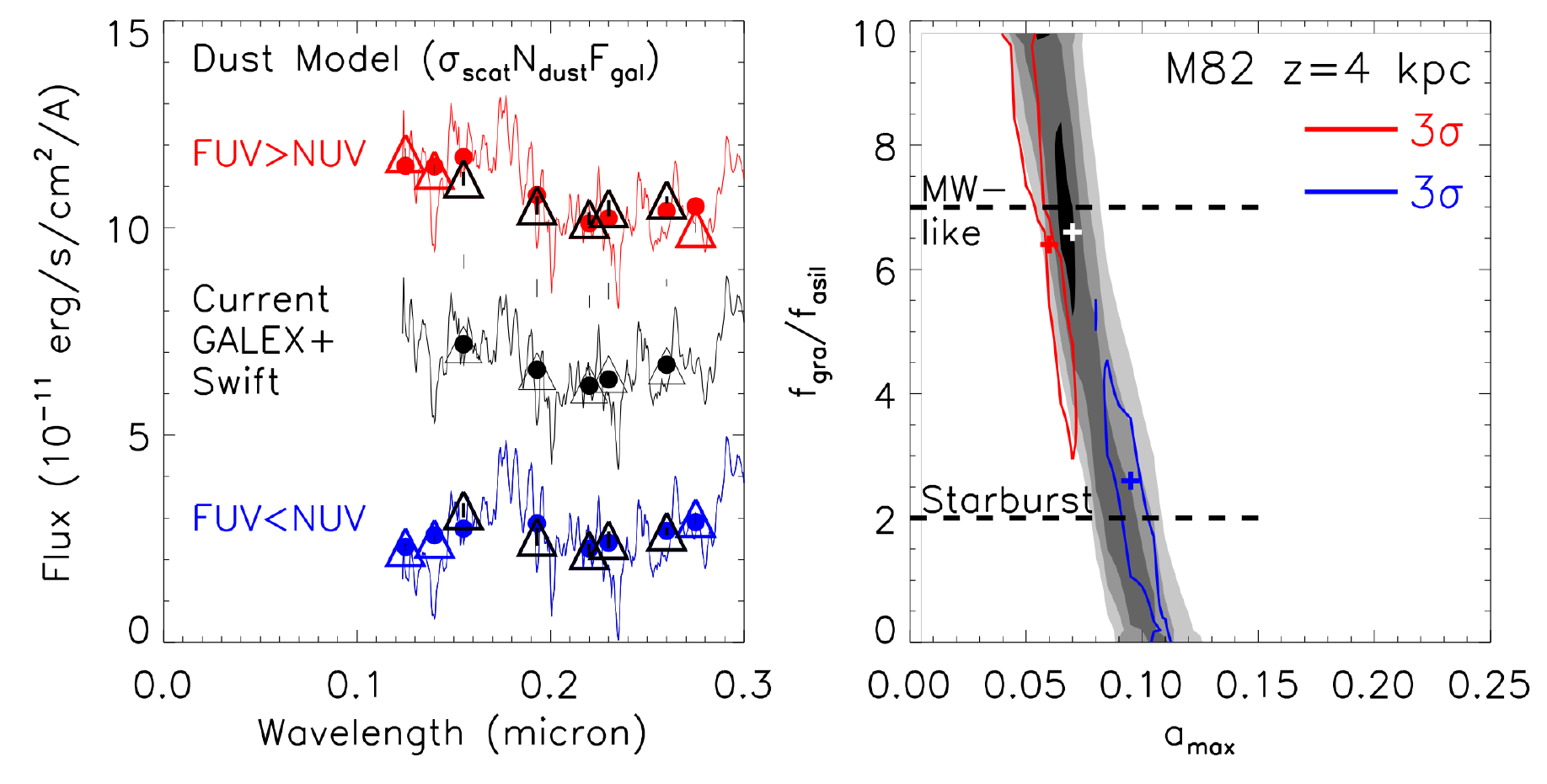}
\caption
{\small \textbf{Left:} Scattering illuminates the complex, dusty wind in M82. Since $\tau_{\text{halo}} \ll 1$, the incident spectrum is height invariant and one can map the dust properties. \textbf{Center:} $\sigma_{\lambda,\text{scat}}$ is determined by fitting the data with a dust model. The black points are measured at $z=4$~kpc above the midplane of M82, and the red and blue points (offset) show potential SEDs with new data. \textbf{Right:} In a silicate-graphite model, variations in $\sigma_{\lambda,\text{scat}}$ are due to changes in dust composition ($y$-axis) or grain size ($x$-axis).  Existing data do not constrain the composition (1, 2, 3, and 4$\sigma$ contours in grayscale), but new data (red and blue 3$\sigma$ contours) would do so.
\label{figure:m82}
}
\end{figure}

\paragraph{\sf Scattering:} CGD near the disk scatters escaping starlight, forming a reflection nebula (RN). RNs can be detected around highly inclined galaxies, where the extraplanar light can be cleanly separated from the disk and the strong forward-scattering is mitigated by integrating through the halo. 
RNs are easiest to detect in the UV band (Figure~\ref{figure:m82}), where the sky is dark and the scattering cross-section ($\sigma_{\text{scat}}$) is high \citep{Hoopes2005,Seon2014,Shinn2015,HK2014}, and stacked \textit{GALEX} data show that RNs exist around most large, edge-on, star-forming galaxies within 100~Mpc (Figure~\ref{figure:stacks}). Initial work based on radiative transfer models finds $10^6$-$10^7 M_{\odot}$ of CGD within $\sim$10~kpc of the disk \citep{HK2016,Baes2016}.  This makes RNs a very promising probe of CGD around normal galaxies.

RNs constrain the CGD properties: the difference between the galaxy spectral energy distribution (SED), corrected for extinction through the disk, and the halo SED determines $\sigma_{\lambda,\text{scat}}$, which depends on the dust properties. In the silicate-graphite model that accurately describes extinction along Galactic sightlines \citep{Mathis1977,Nozawa2013}, $\sigma_{\text{scat}}$ varies with the proportion of silicate grains and the maximum grain size. The observational requirement is a SED covering 1200-2800\AA. Most RN work is based on archival data, and only a few galaxies have sufficient data to make this measurement. Most are missing the required FUV ($\lambda < 2000$\AA) data, and Figure~\ref{figure:m82} shows the constraints on dust properties that can be achieved with additional data. 

Acquiring a sample for population studies requires measuring the SEDs for hundreds of RNs. This maps to requirements on the angular resolution ($\lesssim$1\arcsec), wavelength coverage (1200-2800\AA), and field of view ($\geq$10\arcmin\ to cover the halo). Efficiently making such measurements requires a large collecting area. The existing FUV instruments (\textit{HST} Advanced Camera for Surveys/Solar Blind Channel and the \textit{Astrosat} Ultraviolet Imaging Telescope) either have too small a field of view (\textit{HST}) or too little collecting area (\textit{Astrosat}). However, these requirements are met by CETUS, which has a 17\arcmin field of view, point sensitivity of 26-27~mag, and $\theta < 0.55$\arcsec. 

\medskip
\noindent \textbf{\sf \large 4. Galactic Winds}
\medskip

The most spectacular (and perhaps most important) dusty outflows are in galactic superwinds. The exemplar is M82 (Figure~\ref{figure:m82}), for which the dust is visible in absorption, emission \citep{Engelbracht2006,Roussel2010,Yamagishi2012,Contursi2013,Hutton2014}, and scattered light \citep{Hoopes2005,Hutton2014}. Both PAHs and larger grains are seen in nearby winds with the \textit{Spitzer} IRAC and \textit{Herschel} PACS and SPIRE cameras \cite{McCormick2013,Melendez2015,McCormick2018}. Dusty winds are ubiquitous and contain as much as 20\% of the dust in the galaxy disk, suggesting that episodic outflows are more effective than ongoing, low-level supernova feedback at removing dust. 

However, spectro-polarimetry of optical emission lines in M82 reveals that the dust is kinematically decoupled from other, faster phases and that most dust will not escape the galaxy \citep{Yoshida2011}. At the same time, high-resolution FIR measurements reveal ``worms'' connected to H$\alpha$ filaments \citep{Melendez2015} that indicate a connection between the cool and warm ionized gas. The dust properties may also depend on the primary driving mechanism \citep{Zhang2018}; in quasar outflows, radiation pressure may fully expel dust from galaxies \citep{Murray2005,Ishibashi2016} where hot gas fails to do so. Finally, winds can alter the dust: in at least one galaxy, NGC~891, \citet{Bocchio2016} report a vertical grain size gradient that indicates shock processing.

Dust in winds remains poorly studied except in the most famous winds. A census of dust masses will require observations with existing observatories, as well as new capabilities. Winds can be studied using the same techniques described in Section~2. To resolve wind structures in the FIR beyond 30~Mpc requires much better resolution and sensitivity than \textit{Herschel} (i.e., \textit{Origins}), especially since the starbursts will be very bright, point-like sources. This would also enable dust emissivity maps that constrain the dust origin \citep{Suzuki2018}. The requirements to use RNs to study winds are similar to those for normal galaxies, and so are met by CETUS, which also has a long-slit spectrometer that can separate dust and ionized gas in bright winds. Spectro-polarimetry of filaments should also be possible within 100~Mpc with large ground-based telescopes.

Masses can also be measured from unresolved dusty outflows. For example, the optical--FIR SEDs of AGNs show a separate emission peak for the dust component \citep{Baron2019}. By simultaneously constraining the AGN luminosity and peak location, the mass and covering fraction of the dusty wind are found. This technique can be applied to many AGNs in a deep survey, but we emphasize that understanding the nearby CGD is important to interpreting distant outflows. 

\medskip
\noindent \textbf{\sf \large 5. Summary of Requirements}
\medskip

Emission, extinction/absorption, and scattering provide independent insights, and the combination is very powerful. Here we list requirements to observe the CGD in a large sample. The most progress can be made with a large FIR telescope and a medium-sized, wide field UV telescope.

\bigskip
\hspace{-0.5cm}
\begin{tabular}{ |p{2in}|p{2in}|p{1.8in}|  }
\hline
\rowcolor{tablealt} \textbf{\sf Detection Technique} & \textbf{\sf Methodology} & \textbf{\sf Required Capabilities} \\
\hline
\rowcolor{red!15!white}
Emission \newline (mid-IR to sub-mm) &
Fit SED to find temperature and size. Flux~$\propto$~mass. 
Need highly inclined galaxies within $d<60-100$~Mpc.
&
\specialcell[t]{
$\theta$: 2\arcsec-4\arcsec @ 160~$\mu$m \\ Sens./hr: 1~mJy @ 160~$\mu$m \\
FoV: $\geq$10\arcmin
}
\\
\hline
\rowcolor{green!15!white} Extinction \newline (optical to near-IR) &
Measure reddening around galaxies at $z\gtrsim0.1$ in stacks. $\tau_{\lambda}$ gives composition, mass.  &
\specialcell[t]{Sky Area: $\geq $10,000~deg$^2$ \\ Sensitivity: $>$23~mag }
\\
\hline
\rowcolor{violet!20!white} Scattering \newline (UV) & 
Compare RN, galaxy SED to find $\sigma_{\lambda,\text{scat}}$. Flux~$\propto$~mass. Need highly inclined galaxies within $d<100-200$~Mpc.
& 
\specialcell[t]{
$\theta$: 1\arcsec\  \\ Sensitivity: 26~mag  \\ $\lambda$: 1200-2800\AA\ (filters) \\ FoV: $\geq$10\arcmin
}
\\
\rowcolor{cyan!20!white} Depletion \newline (UV) & 
Measure depletion ratios from CGM absorption lines
& 
\specialcell[t]{
$A_{\text{eff}}$: $>$1,000~cm$^2$ \\ $R$: $\geq$20,000 (NUV) \\ $\lambda$: 1200-3000\AA\
}
\\
\hline
\end{tabular}

\clearpage
\noindent {\bf \large References}

\renewcommand{\section}[2]{}
{  

}

\end{document}